\documentclass[twoside]{article}

\usepackage{PRIMEarxiv}

\usepackage[utf8]{inputenc} 
\usepackage[T1]{fontenc}    
\usepackage{hyperref}       
\usepackage{url}            
\usepackage{booktabs}       
\usepackage{amsfonts}       
\usepackage{nicefrac}       
\usepackage{microtype}      
\usepackage{lipsum}
\usepackage{fancyhdr}       
\usepackage{graphicx}       
\graphicspath{{media/}}     

\title{Lung Cancer Classification from CT Images Using ResNet
}

\author{
  Olajumoke O. Adekunle \\
  Department of Computer Science, \\
  University of Ibadan, \\
  Ibadan, Nigeria\\
  \texttt{jumadek101@gmail.com} \\
   \And
  Joseph D. Akinyemi \\
  Department of Computer Science, \\
  University of York, \\
  York, United Kingdom\\
  \texttt{joseph.akinyemi@york.ac.uk} \\
  \And 
  Khadijat T. Ladoja, Olufade F.W. Onifade \\
  Department of Computer Science, \\
  University of Ibadan, \\
  Ibadan, Nigeria\\
  \texttt{\{kt.bamigbade, ofw.onifade\}@ui.edu.ng} \\
}

\begin{document}
\maketitle

\begin{abstract}
Lung cancer, a malignancy originating in lung tissues, is commonly diagnosed and classified using medical imaging techniques, particularly computed tomography (CT). Despite the integration of machine learning and deep learning methods, the predictive efficacy of automated systems for lung cancer classification from CT images remains below the desired threshold for clinical adoption. Existing research predominantly focuses on binary classification, distinguishing between malignant and benign lung nodules. In this study, a novel deep learning-based approach is introduced, aimed at an improved multi-class classification, discerning various subtypes of lung cancer from CT images. Leveraging a pre-trained ResNet model, lung tissue images were classified into three distinct classes, two of which denote malignancy and one benign. Employing a dataset comprising 15,000 lung CT images sourced from the LC25000 histopathological images, the ResNet50 model was trained on 10,200 images, validated on 2,550 images, and tested on the remaining 2,250 images. Through the incorporation of custom layers atop the ResNet architecture and meticulous hyperparameter fine-tuning, a remarkable test accuracy of 98.8\% was recorded. This represents a notable enhancement over the performance of prior models on the same dataset.
\end{abstract}


\section{Introduction}
Cancer is a global menace, claiming 9.6 million lives in 2018. In 2020, 2.2 million cases were diagnosed, resulting in 1.8 million deaths worldwide \cite{Sharma2022}. Predominant forms include rectal, colon, lung, prostate, breast, and skin cancers. Among these, lung cancer stands as a leading contributor to worldwide cancer-related fatalities. Lung carcinomas arise from pathological changes in epithelial cells, leading to malignant growth. This unchecked proliferation can invade surrounding tissues or distant areas of the body. Non-small-cell lung Carcinoma (NSCLC) poses less threat, while Small-cell Lung Carcinoma (SCLC) spreads rapidly \cite{Gao2018}. Symptoms include hemoptysis, weight loss, dyspnea, and chest pains.

Timely detection significantly impacts prognosis and survival rates, with Computed Tomography (CT) playing a pivotal role in diagnosis. This paper explores the application of ResNet-50, a deep learning model, to enhance the precision of lung cancer classification from CT scans due to its alarming global fatality rate. Research shows that 85\% of cases occur in heavy or chain smokers, while 10-15\% affect non-smokers \cite{Miglio2021}. Lung cancer can be detected through X-rays or CT scans.

Radiologists face the challenge of distinguishing benign from potentially malignant lung nodules. Early diagnosis is critical, necessitating accurate predictive algorithms. Deep learning models offer a personalized approach to assessing lung cancer risk from CT scans. Convolutional Neural Networks (CNN) have proven effective in object recognition, showing promise in medical image analysis. While deep learning techniques have achieved impressive classification accuracies, there is room for improved performance, especially in sensitive medical diagnoses.

This work adapts a pre-trained ResNet model for lung cancer prediction using 15,000 histopathological lung tissue images from the LC25000 dataset \cite{Borkowski2019}. By extending the ResNet model and fine-tuning parameters, we achieve improved performance compared to existing methods on lung cancer prediction using the LC25000 dataset.
The rest of this paper is organized as follows; Section 2 presents a review of previous works in lung cancer prediction from CT images, Section 3 presents the methodology adopted in this research, Section 4 describes the experiments and discusses the results while Section 5 concludes the paper along with recommendations and future work.

\section{Related Works}
In this section, a comprehensive review and analysis of the current body of empirical research concerning the classification of lung cancer from CT images is presented. This analysis has enabled us to discern the primary challenges of lung cancer detection and classification. These identified challenges form the bedrock upon which our multiclass classification model is built.

\subsection{Cancer Classification and Detection}
Lung cancer has emerged as a major health threat in recent years, with patient survival largely depending on the stage at which the disease is diagnosed. Early detection significantly improves survival rates. To aid radiologists, several computer-aided detection (CAD) systems have been developed for identifying lung nodules. With the success of deep learning in image classification, researchers have increasingly applied these techniques to medical imaging, including lung nodule detection. \cite{nakrani2020resnet} proposed a lung nodule detection method using ResNet on CT scan images. The approach involves two main stages: pre-processing and nodule detection. Morphological operations are used to segment the lungs, followed by a convolutional neural network to detect nodules. Designed to serve as a supportive tool for radiologists, the system aims to reduce their workload. The model was tested using the LIDC dataset, which includes CT images from 1,010 patients, achieving a top-5 accuracy of 95.24\%.

\cite{Shafi2022} introduced an innovative cancer diagnostic model amalgamating deep learning with a support vector machine (SVM). This model was engineered to discern physiological and pathological changes in soft tissue cross-sections, particularly those related to lung cancer lesions. The initial training phase involved recognizing lung cancer by measuring and comparing specific profile values in CT images from patients and control subjects. Subsequently, the model underwent testing and validation using CT scans from a separate set of patients and control subjects. Analyzing a dataset comprising 888 annotated CT scans from the publicly available LIDC/IDRI database, the proposed model attained an impressive accuracy rate of 94.00\% in detecting pulmonary nodules, a critical indicator of early-stage lung cancer.

 \cite{Jain2022} highlighted the significance of feature extraction through Kernel Principal Component Analysis (KPCA) integrated within the CNN architecture. This emphasizes the importance of identifying relevant features in histopathological images for accurate cancer detection.

ResNet-50 is a convolutional neural network architecture that has shown exceptional performance in various computer vision tasks, including image classification. Many researchers have utilized the ResNet-50 model or modified versions of it to classify lung cancer from CT images. 

\cite{Mangal2020} proposed an innovative computer-aided diagnostic system leveraging convolutional neural networks (CNNs) for the diagnosis of squamous cell carcinomas and adenocarcinomas in the lung, as well as adenocarcinomas in the colon. Their study incorporated a dataset comprising 2500 digital images sourced from the LC25000 dataset. For lung cancer classification, a relatively straightforward neural network architecture was employed, effectively categorizing histopathological slides into squamous cell carcinomas, adenocarcinomas, and benign cases. Similarly, in colon pathology, the model distinguished between adenocarcinomas and benign cases. Impressively, the diagnostic accuracy achieved surpassed 97.00\% for lung cancer and exceeded 96.00\% for colon cancer.

\cite{Kumar2024} introduces a support system built on ResNet-50, EfficientNet-B3, and ResNet-101, using transfer learning to enhance prediction accuracy. Utilizing 1,000 DICOM images from the LIDC-IDRI repository, the models classify lung cancer into four categories, with the Fusion Model achieving 100\% precision for Squamous Cell classification, the Fusion Model and ResNet-50 achieved a precision of 90\% To improve generalization and reduce overfitting, data augmentation techniques were applied.

\subsection{Lung Cancer Classification}
Several studies collectively highlight the efficacy of Convolutional Neural Networks (CNNs) as a pivotal tool in the detection of lung cancer. \cite{Hatuwal2020} introduced a method utilizing CNNs for the analysis of histopathological images, successfully categorizing them into three distinct classes: benign, Adenocarcinoma, and squamous cell carcinoma. Their model demonstrated remarkable training and validation accuracies of 96.11\% and 97.20\%, respectively. Taking a slightly different approach, \cite{Liang2020} enhanced the VGG19 architecture for the classification of lung cancer biopsy images. Incorporating advanced augmentation techniques, their fine-tuned VGG19 model achieved an impressive accuracy rate of up to 97.73\%. This adaptation of existing models showcases the potential for refining established architectures to attain superior performance in lung cancer classification.

\cite{Kalaivani2020} focused on detecting lung cancer from CT images, employing a densely connected convolutional neural network (DenseNet) in conjunction with an adaptive boosting algorithm. Through this approach, they classified lung images into normal or malignant categories. With a dataset comprising 201 lung images, their proposed method achieved a commendable accuracy of 90.85\%. This underscores the importance of data partitioning for effective training and evaluation of machine learning models.

\cite{Akinyemi2023} introduced a novel approach utilizing deep learning to identify lung and colon cancer. They employed the EfficientNet-B7 architecture and developed a model that utilized the LC25000 dataset for training. To enhance the model, they integrated four additional dense layers into the pre-existing architecture. This modified architecture was then trained on 224×224×3 dimension images. The ultimate dense layer was responsible for classifying input images into one of five categories: Lung benign tissue, Lung adenocarcinoma, Lung squamous cell carcinoma, Colon adenocarcinoma, and Colon benign tissue. The training process encompassed 18,000 images from the dataset, while validation involved 2,000 images and testing employed a separate set of 5,000 images. Over the course of 44 epochs, the model achieved impressive results, boasting detection accuracy rates of 99.63\% during training, 99.50\% during validation, and 99.7\% during testing. In comparison to existing models utilizing the LC25000 dataset, their proposed model displayed remarkable superiority, thereby validating its effectiveness.

In the pursuit of accurate lung and colon cancer identification through histopathological images, \cite{Garg2021} harnessed the power of pre-trained convolutional neural network (CNN)-based models, incorporating enhanced augmentation techniques. They undertook an extensive training process with eight distinct pre-trained CNN models, including VGG16, NASNetMobile, InceptionV3, InceptionResNetV2, ResNet50, Xception, MobileNet, and DenseNet169, using the LC25000 dataset. The evaluation of these models was based on precision, recall, F1 score, accuracy, and AUROC score. The results demonstrated exceptional performance, with all eight models achieving accuracy rates above 96.00\%.

The studies conducted by \cite{Mangal2020} \cite{Garg2021}, \cite{Masud2021} and \cite{Shafi2022} collectively illuminate key insights in the realm of lung and colon cancer detection. \cite{Mangal2020} and \cite{Garg2021} underscored the efficiency of Convolutional Neural Networks (CNNs) in accurately classifying histopathological images for lung and colon cancer diagnosis. Their research showcases the robustness of CNNs as a pivotal tool in the field of medical image analysis. These studies employ a diverse range of techniques and models, including ensemble learning, Kernel Principal Component Analysis (KPCA), and deep learning architectures. This diversity in approaches reflects the adaptive nature of methods in addressing the complex challenge of lung cancer detection.

Remarkably high accuracy rates are reported across all studies, demonstrating the remarkable potential of these models. \cite{Mangal2020} achieved accuracy rates exceeding 97.00\% for lung cancer and surpassing 96.00\% for colon cancer. The work of \cite{Garg2021} exhibits exceptional performance, with accuracy ranging from 96.00\% to 100\% across eight pre-trained CNN models. \cite{Masud2021} achieved an accuracy rate of up to 96.33
\% in distinguishing between different types of lung and colon tissues.

\cite{tawfeek2025enhancing} stated that lung cancer remains the deadliest cancer, prompting the development of a novel Lung Cancer Risk Prediction (LCRP) model to detect and classify pulmonary diseases like pneumonia, tuberculosis, and lung cancer using X-ray and CT images. The LCRP framework comprises four key modules: data preprocessing, data augmentation, image segmentation, and prediction. It integrates three deep learning models—sequential, functional, and transfer models—alongside CNNs and uses Mask R-CNN for improved image segmentation and classification. The model employs dual optimizers (Adam and SGD) to boost training efficiency. LCRP demonstrates superior performance, achieving 98.5\% accuracy while reducing training complexity and computational costs.

These studies showcase adaptability by utilizing a diverse range of datasets, including LC25000 \cite{Borkowski2019}, LZ2500 \cite{Jain2022}, NLST \cite{Jain2022}, and NCI Genomic datasets \cite{Jain2022}. This versatility in data sources enhances the potential applicability of the models across different clinical scenarios. In evaluating their models, these studies employed a comprehensive set of metrics including precision, recall, F1-score, and Area Under the Curve (AUC), providing a thorough assessment of model performance.

The studies on lung cancer detection using machine learning and deep learning methods exhibit notable promise, but they are not without potential limitations. These include the reliance on specific datasets that may not fully represent clinical diversity, potential issues with sample size and class distribution imbalance, the impact of data preprocessing and augmentation techniques on model performance, varying image quality in histopathological slides, challenges in interpreting features extracted by deep learning models, computational resource demands for training and deployment, risks of overfitting and the need for robust generalization, the necessity for clinical validation in real-world settings, ethical and legal considerations in healthcare applications, and the importance of long-term follow-up data for assessing patient outcomes. Addressing these concerns will be essential for ensuring the reliability, effectiveness, and ethical deployment of these AI-based tools in clinical practice. This work carefully approaches the task of lung cancer detection in CT images in order to combat some of these problems such as overfitting, model explainability and data imbalance.

\section{Methodology}
\label{sec:method}
The early detection of lung cancer is known to be of great significance in the effective treatment and survival of the disease. This research proposes a deep-learning model for the classification of cancerous and non-cancerous cells based on CT Images of lung tissues. This section presents a description of the materials and methods employed in this research. The classification of lung CT images was realized by using transfer learning techniques on a relatively large dataset of lung CT images. The generic process involved loading images in batches from the dataset, preprocessing the images, training the deep learning model on the images, validating the deep learning model on a separate subset, fine-tuning the model and repeating the process until a satisfactory classification accuracy was obtained. For a proper evaluation and to prevent overfitting, the dataset was split into three subsets (training, validation and test sets). The validation set was used to evaluate training performance, and only when the model's performance on the validation set was satisfactory would the model be tested on the test set.

\subsection{The Dataset}
\begin{figure}[h]
    \centering
    \includegraphics[width = .5\textwidth]{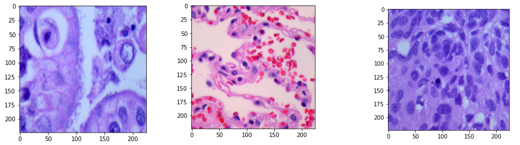}
    \caption{ Image samples in the dataset.}
    \label{fig:fig1}
\end{figure}

The dataset used was the Lung and Colon Histopathological Image Dataset (LC25000) \cite{Borkowski2019}. 5 classes of 5,000 images each make up the 25,000 colour images of lung and colon tissues in the LC25000 dataset. For this work, only the 15,000 images of lung tissues in LC25000 were used. The 5 classes in the dataset are \textit{Lung benign tissue}, \textit{Lung adenocarcinoma}, \textit{Lung squamous cell carcinoma}, \textit{Colon adenocarcinoma}, and \textit{Colon benign tissue}; only the three classes corresponding to the lung images were used for this work. The 15,000 images of lung tissues were split into 10,200 training set images, 2,550 validation set images and 2,250 test set images. Samples images in the dataset for each of the three lung tissue classes are shown in Fig. ~\ref{fig:fig1}.

Data preprocessing involves converting raw data into a format usable for a machine learning model. In our approach, input images were resized to a 224 by 224 pixels size to make them compatible with the deep learning model used for feature extraction and image classification. The images in the dataset are all high-resolution histopathological images, so there was no need for more preprocessing than resizing the images to suit the input dimensions of the deep learning model employed.

\subsection{The Deep Learning Model}
In this work, a pre-trained deep learning model was employed for feature representation and classification of lung CT images. We chose the ResNet-50 \cite{He2016} architecture because of its skipped connections, which enable it to combat the vanishing gradient problem and model overfitting while training through a very deep network. These features make it possible to learn robust feature representations, which in turn enhance efficient classification. To better adapt the ResNet model to the task of lung cancer classification, we removed the final classification layer and added custom layers on top of the architecture. ResNet was originally trained for object detection on the large ImageNet dataset \cite{Deng2009} containing 1000 unique object classes. Given that the ResNet architecture is quite deep and contains several helpful layers, including dropout layers (used to combat overfitting), we only needed to add a few more layers to adapt the learnt features to our own classification task. Table ~\ref{tab:tab1} shows an overview of the resulting architecture after adaptation for the lung cancer classification task.

\begin{table}
    \caption{ An overview of the adapted ResNet architecture (showing custom layers)}
    \centering
    \begin{tabular}{lll}
    \toprule
    Layer (type) &	Output shape &	Parameters\\
    \midrule
    Lambda (Lambda) &	(None, 224, 224, 3) &	0\\
    
    ResNet50v2 (Functional) &	(None, 2048) &	23564800 \\
    Dense (Dense) &	(None, 128) &	262272 \\
    Dense\_1 (Dense) &	(None, 3) &	387 \\
    \bottomrule
    \end{tabular}
    \label{tab:tab1}
\end{table}

As shown in Table ~\ref{tab:tab1}, the model takes an input image of (None, 224, 224, 3) which means that it accepts images with a size of 224 × 224 pixels and 3 colour channels (Red, Green Blue), while the first dimension, None stands for the batch size. The next layer is the actual pre-trained ResNet architecture with the final output layer removed. This architecture thus returns a vector of 2048 values (as a result of the flattening layer just before the final output layer). The third layer is the Dense layer, also known as the fully connected layer, which receives the 2048-dimensional vector from the ResNet flatten layer and produces a 128-dimensional vector as its output. The last layer is named Dense\_1 (just to differentiate it from the previous Dense layer) with an output shape of (None, 3), which means that it produces a 3-dimensional vector, which corresponds to the number of classes in our classification task.

\begin{figure}[h]
    \centering
    \includegraphics[width = .5\textwidth]{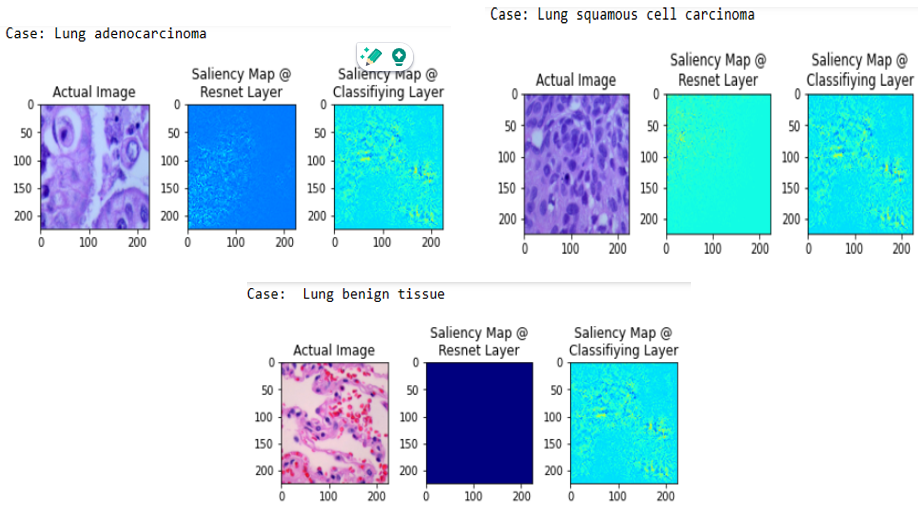}
    \caption{Saliency maps showing some of the features extracted by the deep learning model from CT images of Top-left (\textit{Lung adenocarcinoma}); Top-right (\textit{Lung squamous cell carcinoma}); bottom (\textit{lung benign}).}
    \label{fig:fig2}
\end{figure}

Investigating the features extracted and learned by a deep learning model can be interesting. The ResNet model utilizes the learned features from its intermediate layers, which are also referred to as "feature maps." These feature maps capture a hierarchical representation of the input image, gradually extracting features that become more intricate and abstract. Typically, the layers closer to the input capture low-level features like edges, textures, and basic shapes, while deeper layers capture more complex and abstract visual representations. Fig. ~\ref{fig:fig2} shows the saliency maps obtained from the pre-trained ResNet architecture and those obtained from the final classification layer.

As seen from the extracted features in Fig. ~\ref{fig:fig2}, the actual image represents the input image that has been resized and augmented before the pre-trained model performs the feature extraction. The rest of the images show the feature maps as extracted by the pre-trained ResNet architecture and the final classification layer.  However, we see high similarity in the feature patterns extracted in the classification layer. In fact, the maps look identical at this final classification layer, but a closer look shows that the feature pattern of the benign tissues is markedly different from those of the two malignant tissues. This enables the architecture to accurately differentiate between the benign and malignant tissues as seen later in our results.

\section{Results and Discussion}
\label{sec:results}
We conducted all experiments on Google Colab and Kaggle, utilizing Python 3.5.3 with a GPU configuration of P100, 15GB of RAM, and a 500GB HDD. Our implementation incorporated essential libraries, including TensorFlow for deep learning model development and validation, Scikit-learn for performance assessment, Matplotlib for visualization, and NumPy for array manipulation. The experiment was carried out using the adapted ResNet-50 model trained for 18 epochs (set to train for 25 epochs but stopped at 18 epochs to prevent overfitting) with a batch size of 25 (the image data was loaded in a batch of 25) and a learning rate of 0.001.

To evaluate the performance of the developed lung cancer classification model, 4 standard classification metrics were used: accuracy, precision, recall and F1-score. Following the standard convention, the F1-score was chosen as the evaluation metric on which the final performance of the model is based.

\begin{figure}[h]
    \centering
    \includegraphics[width = .5\textwidth]{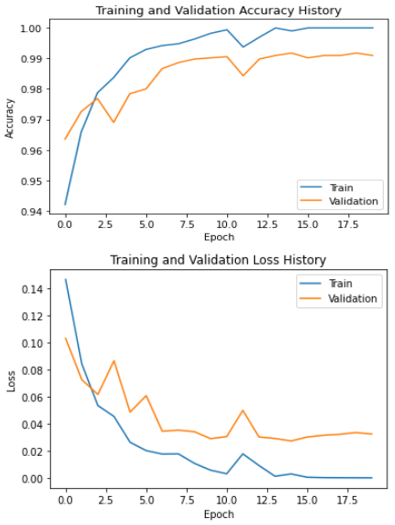}
    \caption{Training and validation curves of the lung cancer prediction model. Top: accuracy curves. Bottom: Loss curves.}
    \label{fig:fig3}
\end{figure}

A validation accuracy of 0.99 (or 99\%) means that the model is correctly classifying 99\% of the examples in the validation dataset. A training accuracy of 1.00 (or 100\%) means that the model correctly classifies all the examples in the training dataset. As shown in Fig.~\ref{fig:fig3}, a training loss of less than 0.02 means that, on average, the predictions made by the model are very close to the true labels in the training data. Lower training loss values generally indicate that the model is learning and fitting well to the training data.  A validation loss of less than 0.04 suggests that the model's predictions are close to the true labels in the validation set. Therefore, judging from the graph, overfitting is minimal. As can be seen in Fig.~\ref{fig:fig3}, the validation performance keeps in step with the training performance (both for accuracy and loss), so overfitting is minimal, and we can conclude that the model achieved good generalization.

\begin{table}
    \caption{Evaluation metrics scores of the lung cancer prediction model}
    \centering
    \begin{tabular}{lllll}
    \toprule
Classes & Precision & Recall & F1 -score & samples \\
\midrule
Benign & 1.0000 & 1.0000 & 1.0000 &  750 \\
Adenocarc. & 0.9775 & 0.9867 & 0.9821 & 750 \\
Squamous & 0.9865 & 0.9773 & 0.9819 & 750 \\
Accuracy & & & 0.9880 & 2250 \\
Macro avg &	0.9880 & 0.9880 & 0.9880 & 2250 \\
Weighted avg & 0.9880 & 0.9880 & 0.9880 & 2250 \\
\bottomrule
    \end{tabular}
    \label{tab:tab2}
\end{table}

\subsection{Model Evaluation}
\begin{figure}[h]
    \centering
    \includegraphics[width = .5\textwidth]{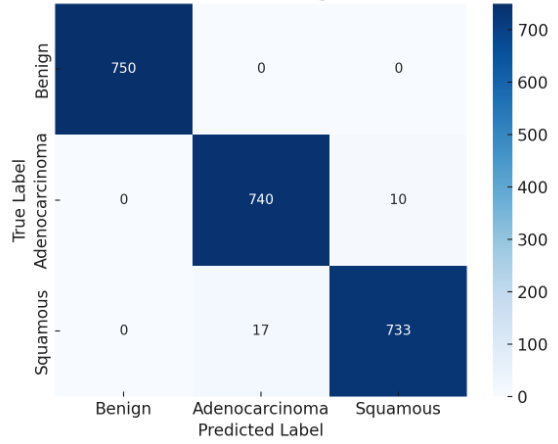}
    \caption{Confusion matrix of the lung cancer prediction model.}
    \label{fig:fig4}
\end{figure}

Model evaluation helps us understand how well the deep learning model performs on real-world data. It gives insights into how accurately the model can make predictions or classifications on previously unseen data. The developed lung cancer prediction model was evaluated using classification accuracy, recall, precision and F1-score on the test dataset containing 2250 CT images. Table 2 shows the performance evaluation of the model via the scores/values of these evaluation metrics on the test set while Fig.~\ref{fig:fig4} shows the confusion matrix. As seen in Table~\ref{tab:tab2}, the benign tissues are perfectly classified with 100\% accuracy because the tissue structure seems markedly different from the malignant classes as seen in the features map of Fig.~\ref{fig:fig2}. This is also reflected in the confusion matrix of Fig.~\ref{fig:fig4}, as all benign tissues in the test set were correctly classified. Of the two malignant classes, it is observed that Squamous cell carcinoma suffers the worst classification performance, having a final F1-score of 0.9819 (or 98.19\%) as seen in Table~\ref{tab:tab2} and a total misclassification of 17 images as seen in Fig.~\ref{fig:fig4}. Again, the misclassifications in the test set are quite restricted between the two malignant classes (as seen in Fig.~\ref{fig:fig4}). This proves the efficiency of deep learning algorithms in correctly classifying benign and malignant tissues. Although the resulting test set is relatively small, the fact that it was separate from the validation set gives confidence that the model had no chance of overfitting the test set during training. Thus the results presented are representative of the generalization ability of the lung cancer prediction model developed in this work.

\begin{table}    
    \caption{ Evaluation metrics scores of the lung cancer prediction model}
    \centering
    \begin{tabular}{llll}
    \toprule
S/N & Author & Model & Accuracy (\%) \\
\midrule
1. & \cite{Garg2021} & ResNet50 & 96.00  \\
2. & \cite{Liang2020} & MFF-CNN & 96.00  \\
3. & \cite{Sikder2021} & CNN & 96.33 \\
4. & \cite{Shafi2022} &	SVM+CNN & 94.00  \\
5. & \cite{Mamun2022} & XGBoost &  94.42  \\
6. & \cite{Jain2022} & FDBNN \& KPCA-CNN & 97.50 \\
7. & \cite{tawfeek2025enhancing} & LCRP & 98.50 \\
8. & \cite{Kumar2024} & Fusion Model \& ResNet-50 & 90.00 \\
9.  & \cite{Sandhya2023} & LungNet & 88.13 \\
10. & \textbf{Our model} & \textbf{ResNet50} & \textbf{98.80} \\
\bottomrule
    \end{tabular}
    \label{tab:tab3}
\end{table}

\subsection{Comparative Analysis}
The performance of the developed model was assessed in comparison with other existing models using the LC25000 dataset. As our work is focused on lung cancer prediction, we compared the performance of our model with those of previous works which were also focused on the lung cancer subset of LC25000 dataset. Table~\ref{tab:tab3} presents a comparative analysis of the performance of the existing models of lung cancer detection in CT images, based on the LC25000 dataset. As seen in Table~\ref{tab:tab3}, our model obtained a reasonably higher classification accuracy than all the models compared. This performance can be attributed to the power of the ResNet50 model in capturing latent visual features that allow it to generalize well to unseen data.

\section{Conclusion}
Lung cancer is a leading cause of death worldwide. As a result, deep learning algorithms are utilized to speed up the essential process of identifying lung cancer and minimizing pathologists' burden. This research presents Lung Cancer Classification from CT Images Using a pre-trained deep learning architecture, ResNet-50. The pre-trained architecture was customized for the task and applied to the 15000 lung CT images in the LC25000 lung histopathological images dataset. The developed model classifies each CT image into one of three classes: benign, adenocarcinoma and squamous cell carcinoma. Each class contains 5000 histopathology images of lung tissues. The developed model achieved training, validation and test accuracies of 100\%, 99\% and 98.80\% respectively showing good generalization of the model to unseen data. Future works could consider investigating larger datasets and possibly more cancer classes and types, as an accurate classification of cancer types could significantly aid diagnosis.
\bibliographystyle{unsrt}  
\bibliography{references}

\end{document}